\documentclass[runningheads]{llncs}
\usepackage{graphicx}
\usepackage{booktabs}
\usepackage{multirow}
\usepackage{amsmath}
\usepackage{graphicx}
\usepackage{subfigure} 
\usepackage{wrapfig}
\usepackage[misc]{ifsym}
\usepackage{color}
\usepackage{hyperref}
\hypersetup{
    colorlinks=true,
    linkcolor=blue,
    filecolor=blue,      
    urlcolor=blue,
    citecolor=blue,
}

\begin{document}\bibliographystyle{splncs04}
\title{Stepwise Feature Fusion: Local Guides Global}

\author{Jinfeng Wang\inst{1,2\footnotemark[1]} \and
Qiming Huang\inst{1\footnotemark[1]} \and
Feilong Tang\inst{1\footnotemark[1]} \and Jia Meng\inst{1} \and Jionglong Su\inst{1\ \textrm{\Letter}} \and Sifan Song\inst{1,2\  \textrm{\Letter}}} 


\institute{Xi’an Jiaotong-Liverpool University, Suzhou, China \and University of Liverpool, Liverpool, UK\\ \email{Jionglong.Su@xjtlu.edu.cn\\
Sifan.Song19@student.xjtlu.edu.cn}\\
\href{https://github.com/Qiming-Huang/ssformer}{https://github.com/Qiming-Huang/ssformer}}
\authorrunning{J. Wang et al.}
\maketitle
\footnotetext[1]{Contributed Equally}

\begin{abstract}
Colonoscopy, currently the most efficient and recognized colon polyp detection technology, is necessary for early screening and prevention of colorectal cancer. However, due to the varying size and complex morphological features of colonic polyps as well as the indistinct boundary between polyps and mucosa, accurate segmentation of polyps is still challenging. Deep learning has become popular for accurate polyp segmentation tasks with excellent results. However, due to the structure of polyps image and the varying shapes of polyps, it is easy for existing deep learning models to overfit the current dataset. As a result, the model may not process unseen colonoscopy data. To address this, we propose a new state-of-the-art model for medical image segmentation, the SSFormer, which uses a pyramid Transformer encoder to improve the generalization ability of models. Specifically, our proposed Progressive Locality Decoder can be adapted to the pyramid Transformer backbone to emphasize local features and restrict attention dispersion. The SSFormer achieves state-of-the-art performance in both learning and generalization assessment. 

\end{abstract}
\section{Introduction}
Colorectal cancer (CRC) is common cancer whose cancer risk may be reduced through early screening and removal of colon polyps \cite{pranet,UACAnet}. However, accurate polyp segmentation is still a challenge due to the variable size and shape of polyps, as well as the indistinct boundaries between polyps and mucosa \cite{pranet}. An accurate segmentation algorithm based on deep learning can effectively improve the accuracy and efficiency of polyp segmentation. Many image segmentation models based on the Convolutional Neural Networks (CNN) recently achieved excellent learning ability in several polyp segmentation benchmarks. \cite{pranet,UACAnet,caranet,srivastava2021msrf,unet++} However, due to the top-down modeling method of the CNN model and the variability in the morphology of polyps but relatively simple structure of the polyps image, this model lacks generalization ability and is difficult to process unseen datasets. To improve the generalization ability of the deep learning model, we shall incorporate the Transformer architecture into the polyp segmentation task.

The Transformer \cite{vaswani2017attention} was initially proposed as a bottom-up model architecture in the natural language processing (NLP) community. Dosovitskiy \emph{et al}. proposed the Vision Transformer (ViT) \cite{dosovitskiy2020image} that achieved superior performance in image classification tasks. The Transformer is different from CNN which the weight parameters are trained in the kernel to extract and mix the features among elements in the receptive field. In contrast, the Transformer obtains similarities of all patch pairs through the dot product between the patch vectors to adaptively extract and mix features between all patches. This enables the Transformer to have an efficient global receptive field and reduces the inductive bias of the model. As a result, the Transformer has a more robust generalization ability than CNN and Multilayer Perceptron-like structures \cite{naseer2021intriguing}. However, the low inductive bias and powerful global receptive field make it difficult for the Transformer model to capture task-specific critical local details adequately. In addition, with the deepening of the Transformer model, the global features are continuously mixed and converged \cite{zhou2021deepvit}, resulting in attention dispersion. These make it difficult for the Transformer model to accurately predict detailed information in the dense prediction task of semantic segmentation. 

In order to achieve high generalization and accurate polyp automatic segmentation, a novel state-of-the-art (SOTA) medical image segmentation model, SSFormer, is proposed which uses a pyramid Transformer encoder \cite{pvtv1,xie2021segformer,pvtv2,simvit} for excellent generalization and multi-scale feature processing capabilities. In our model, the Progressive Locality Decoder (PLD), based on a multi-stage feature aggregation structure, functions as the decoder. The multi-stage feature aggregation structure can enable features of different depths and expressive powers to guide each other, which we believe can address the problems of attention dispersion and underestimation of local features to improve the detail processing ability. Segformer \cite{xie2021segformer} optimized the encoder of the pyramid structure of PVT \cite{pvtv1} and proposed a multi-stage feature aggregation decoder, which predicts features of different scales and depths separately through simple upsampling and then parallel fusion. SETR \cite{zheng2021rethinking} uses the traditional Transformer as the encoder and proposes an MLA decoder with a multi-stage feature aggregation structure. Their excellent performances demonstrate that the decoding method of multi-stage feature aggregation is beneficial to improving the performance of Transformer in dense prediction tasks. Our proposed PLD adopts a stepwise adaptive method to emphasise local features and integrate them into global features, making the fusion of features more efficient.

The main contributions of this paper are: 1) We introduce the pyramid Transformer architecture into the polyp segmentation task to increase the generalization ability of the neural network; 2) We propose a new decoder PLD suitable for Transformer feature pyramids, which can smooth and effectively emphasise the local features in the Transformer to improve the detailed information processing ability of the neural network; 3) Our proposed SSFormer improves the SOTA performances of the ETIS benchmark, CVC-ClinicDB benchmark, and Kvasir benchmark by about 3\%, 1.8\%, and 1\%, respectively.  In addition, SSFormer has achieved state-of-the-art and superior performance in 2018 Data Science Bowl and ISIC-2018 benchmarks.

\begin{figure}[t]

\centering
\subfigure[Overview of SSFormer]{
\includegraphics[width=9cm]{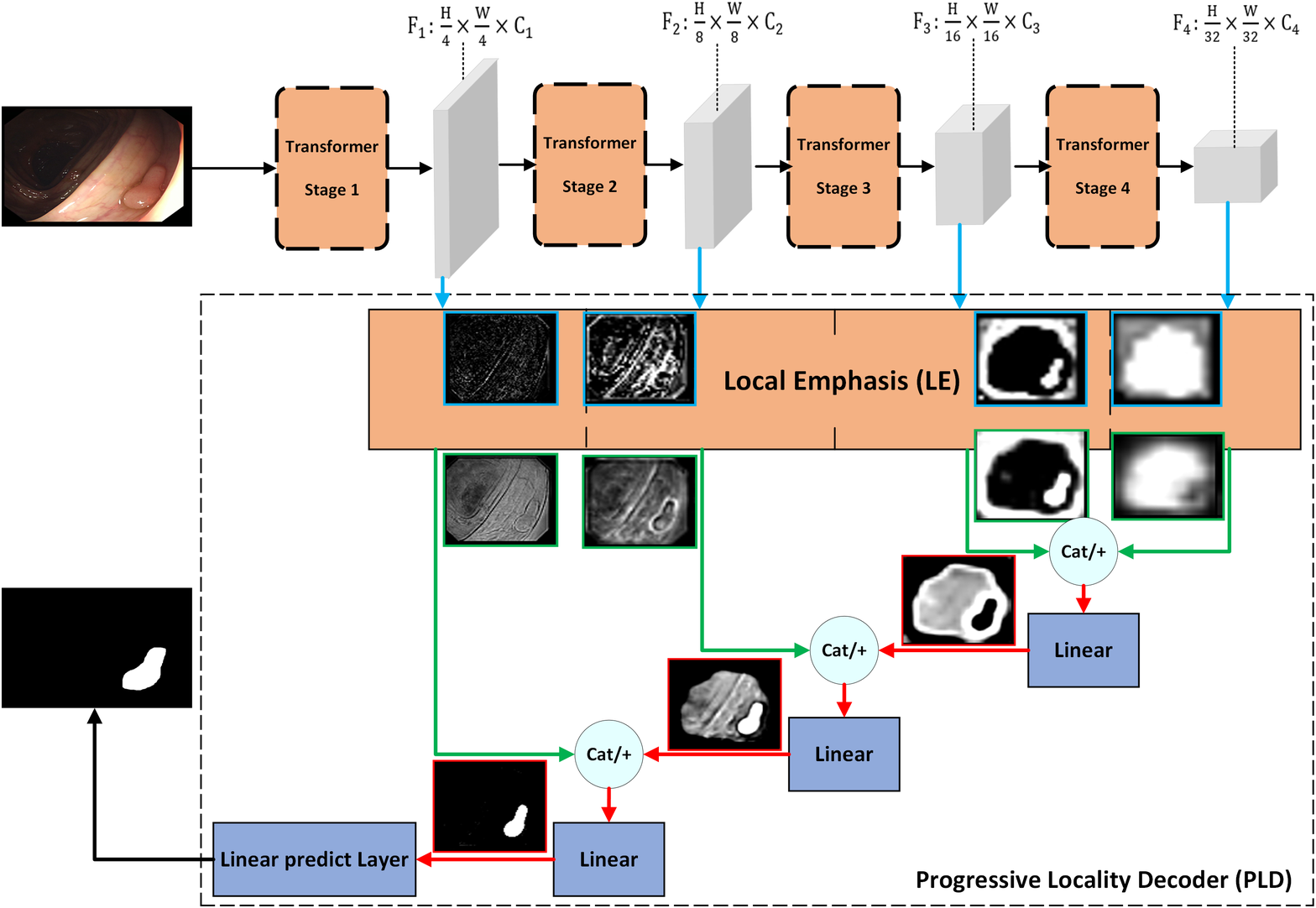}
\label{Overview}
}
\quad
\subfigure[LE]{
\includegraphics[width=1.5cm,height=5cm]{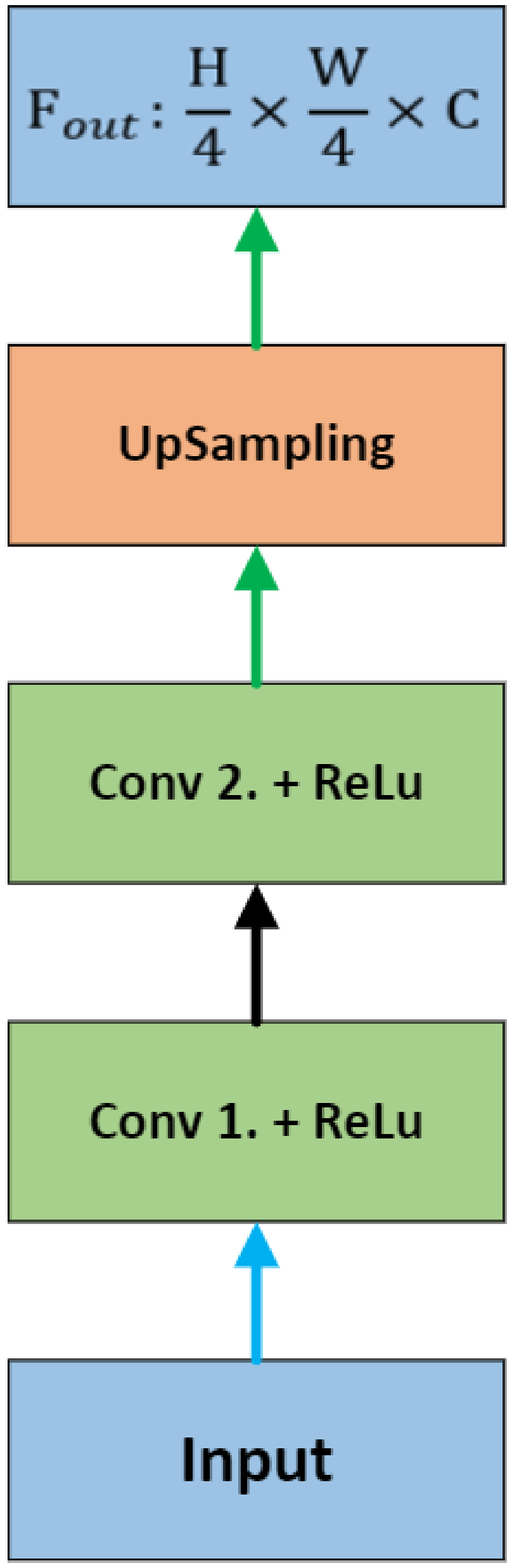}
\label{LE}
}
\caption{(a) the Overview of SSFormer; (b) the structure of Local Emphasis module. In this figure, The lines with arrows and the feature maps next to them represent unemphasized features, local emphasized features, and fused features from top to bottom along the feature stream direction, respectively. The remainder of the PLD in Figure (a), excluding the Local Emphasis (LE), is the Stepwise Feature Aggregation (SFA). Feature fusion units can use concatenation (Cat) or addition (+) operations. }
\end{figure}

\section{Methodology}

\subsection{Transformer encoder}
\label{MiT}
In order for our model to have enough generalization ability and multi-scale feature processing ability to carry out polyp segmentation, we use the Transformer based on the pyramid structure instead of CNN as the encoder. To this end, we adopt the encoder design of PVTv2 \cite{pvtv2} and Segformer to construct the encoder. They both use the convolution operation to replace the PE operation of the traditional Transformer for consistency of spatial information, excellent performance and stability.

\subsection{Aggregate local and global features stepwise (PLD)}
\label{PLD_section}
\begin{figure}[!h]
    \centering
    \subfigure[Attention heatmap for different decoder(SeD is Segformer's Decoder)]{
    \includegraphics[width=12cm]{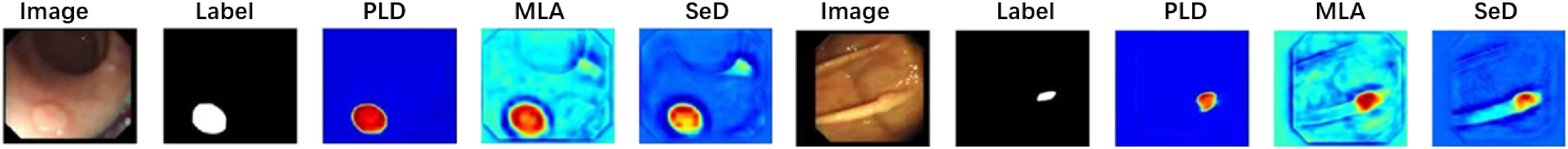}
    \label{AH2}
    }
    \subfigure[Attention Heatmap for the LE module]{
    \includegraphics[width=12cm]{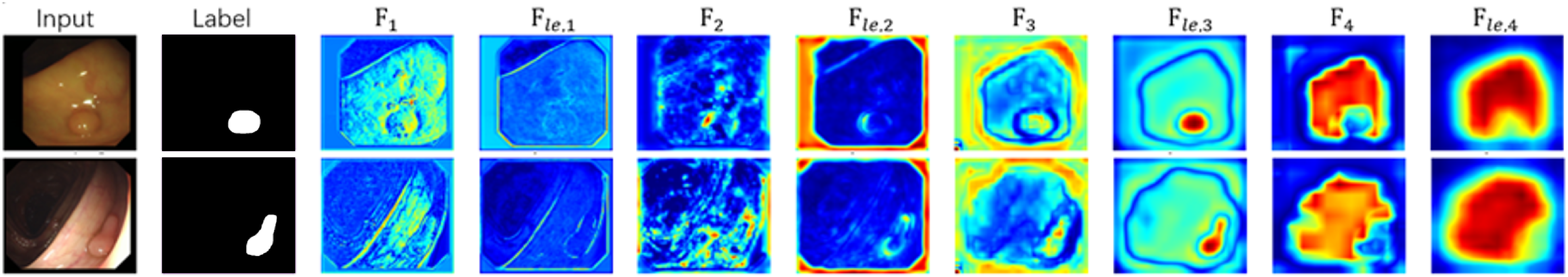}
    \label{AH}
    }
    \subfigure[Attention Heatmap for the SFA]{
    \includegraphics[width=12cm]{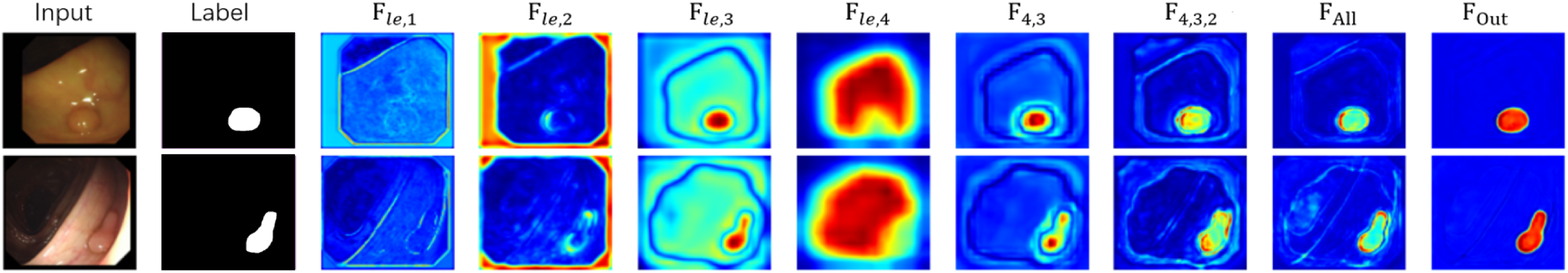}
    \label{AH1}
    }
    \label{attention map}
    \caption{Attention heatmap of feature flow through the PLD process. Figure (a) shows that the LE module successfully focuses the model's attention on critical details. Figure (b) shows that the SFA structure effectively constrains the model's chaotic attention stepwise to fine critical regions.}
\end{figure}

Experiments \cite{raghu2021vision,zheng2021rethinking} have demonstrated that the sufficiency of local features obtained in the shallow part of the Transformer directly affects the performance of the model. However, we believe that the existing Transformer model lacks local and detailed information processing ability to focus on critical detailed features (such as contour, veins and texture). As a result, this makes it difficult for the model to locate the more decisive local feature distribution (mucosa can be considered a distribution composed of local features such as unique veins and textures). We propose a novel multi-stage feature aggregation decoder PLD for feature pyramids to address this issue. Fig. \ref{Overview} shows that the PLD consists of the Local Emphasis (LE) module and the Stepwise Feature Aggregation (SFA) module. The experimental section compares PLD with other existing decoders with various encoders that can generate feature pyramids. We compare the attention distribution before the final prediction of several typical multi-stage feature aggregation decoders for Transformers. As demonstrated in Fig. \ref{AH2}, after PLD fuses multi-stage features, the prediction head can accurately focus on critical targets. In addition, our PLD can be used for other Pyramid Transformer encoders and can improve the model's accuracy. There is a further demonstration in Section \ref{section:AS} .

\subsubsection{Local Emphasis}
In the Transformer, each patch in the image will mix the information of all other patches, even if their correlation is not high. After a large number of self-attention operations, the feature streams will converge, further exacerbating the attention dispersion or attention collapse \cite{zhou2021deepvit}. Furthermore, we argue that the attention matrix in the self-attention mechanism can be viewed as a global non-preset convolution kernel. We designed the LE module using the local receptive field of the convolution kernel to increase the macro weights of the patches around the query patch to refocus attention on neighboring features thus reducing attention dispersion. In Fig. \ref{LE}, the module consists of the convolution operators, activation functions, and a bilinear upsampling layer. We utilize the fixed receptive field of the convolution operator to mix the features of the adjacent patches of each patch, thereby increasing the associated weights of the adjacent patches to the center patch, thus emphasizing the local features of each patch. Since the feature types of the feature streams from different depths are different, we do not share the convolution weights for the feature streams at different levels in the feature pyramid. The formula for strengthening local features is as follows:
\begin{equation}
F_{le,i} = ReLU(Conv_i(C,C)(ReLU(Conv_i(C_{i},C)(F_{i})))),
\end{equation}
where $F_{le,i}$ refer to the local emphasized feature from stage $i$, $Conv_i(C_{in}, C_{out})$ and $Linear(C_{in}, C_{out})$ refer to a convolutional and linear layer with input channel $C_{in}$ and output channel $C_{out}$. From the feature map given in Fig. \ref{Overview}, it can be seen that the LE can effectively clean up cluttered noises and emphasize critical local features. In Fig. \ref{AH}, after the feature stream passes through LE, the disordered attention is re-condensed along with critical details such as contours and boundaries.

\subsubsection{Stepwise Feature Aggregation (SFA)}
 Experiments \cite{raghu2021vision} have demonstrated that the amount of information interacted through residual connections \cite{he2016deep} in the Transformer is more significant than that of the CNN model. This phenomenon can be understood as the weak correlation between the features of different depths in the Transformer, requiring a lot of information interaction for the layers of different depths to guide each other. As such, we believe that direct parallel aggregation of features of different stages with significant differences in depth in Transformer may generate an information gap. 

In order for the feature aggregation to be as smooth as possible, the SFA progressively fuses the features of different levels in the feature pyramid from the top to bottom. From the perspective of the change of feature streams, it can be considered that the local features of the shallower layers are progressively fused into the global features of the deeper layer. This feature fusion method can reduce the information gap between the fused high-dimensional and low-dimensional features. 
As given in Fig. \ref{AH1}, local features gradually guide the attention of the model to critical regions in the SFA. In Fig. \ref{Overview}, the SFA consists of feature fusion units, linear fusion layers, and a linear prediction layer. The feature map of the fused structure in Fig. \ref{Overview} (image with red border) shows that the SFA effectively incorporates local features into high-dimensional features and guides the feature stream into critical regions.
\begin{equation}
\begin{split}
\label{SFA}
F_{i-1,i} = \left \{
\begin{array}{ll}
    Linear(2C,C)(Concat(F_{i-1},F_{i})), \\
    OR, \\
    Linear(C,C)(Add(F_{i-1},F_{i})),
\end{array}
\right.
\end{split}
\end{equation}

Since the feature stream has the same shape after passing through the LE module, we can use concatenation or addition operation in the feature fusion unit as Equation \ref{SFA}. In Table \ref{AS2}, we see that both perform equally well. Concatenation is the default in SSFormer.

\subsection{Stepwise Segmentation Transformer}
Based on the different encoder scales, we propose the SSFormer-S (Standard) and the SSFormer-L (Large) model. They achieve SOTA and competitive performance in several polyp segmentation benchmarks. Details are given in the experimental section. Moreover, SSFormer also achieved SOTA and competitive performance in ISIC-2018 and 2018 DATA Science Bowl. 

\section{Experiments}
\subsection{Experimental Setup}\label{Setup}
\subsubsection{Dataset and Evaluation Matrix}
Since the colon polyp segmentation task requires the model to have both accurate prediction and generalization capabilities, the performance of model on experimental and unseen benchmark datasets needs to be assessed separately. Therefore, following the experimental scheme of MSRF-Net \cite{srivastava2021msrf}, we train and test  SSFormer on the Kavsir-SEG \cite{jha2020kvasir} and CVC-ClinicDB \cite{bernal2015wm} benchmark datasets, respectively, to assess the accurate prediction and learning ability of models in the Kavsir-SEG and CVC-ClinicDB test set, respectively. In order to assess the generalization ability of SSFormer, we tested the model trained in Kavsir-SEG on CVC-ClinicDB and vice versa. 

We refer to the experimental scheme of PraNet \cite{pranet} and UACANet \cite{UACAnet} that randomly extract 1450 images from the Kavsir and CVC-ClinicDB benchmark datasets to construct a training set (For fairness evaluation, we used the same training set as UACANet and PraNet), then test the model trained in this training set on the CVC-ColonDB \cite{bernal2012towards} and ETIS \cite{silva2014toward} benchmark datasets. This test can demonstrate our model's accurate prediction and generalization ability in unseen datasets. Due to the variety of types and sizes of polyps in ETIS, it is the most challenging benchmark. The ISIC-2018 \cite{codella2018skin,tschandl2018ham10000} and 2018 Data Science Bowl \cite{caicedo2019nucleus} benchmark datasets were also used in additional experiments. To unify the performance measures of the above two schemes, we only use mean Dice and mean IoU as evaluation matrices in our experimentation.

\subsubsection{Implementation details}
We implement our model in PyTorch, which an NVIDIA TESLA A100 GPU accelerates. The AdamW optimizer is used with an initial learning rate of 0.0001, a decay rate of 0.1, and a decay period of 40 epochs. The training period is 200 epochs. Our loss function is the combined loss of Dice loss and BCE loss. During training, we resize the image to $352\times352$. We employ random flipping, scaling, rotation, and random dilation and erosion as data augmentation operations. 

\begin{table}[htbp]\scriptsize\centering
\renewcommand\arraystretch{1.2}
\caption{The performance of the SOTA methods was trained and tested on the same benchmark dataset, used to assess learning ability, the scores in the table refer to \cite{srivastava2021msrf}}.
\label{LA}
\begin{tabular}{|l|cc|cc|cc|cc|}
\hline
\multicolumn{1}{|c|}{\textbf{Dataset}} & \multicolumn{2}{c|}{\textbf{CVC-ClinicDB}}             & \multicolumn{2}{c|}{\textbf{Kvasir-SEG}}               & \multicolumn{2}{c|}{\textbf{ISIC-2018}}                & \multicolumn{2}{c|}{\textbf{2018 Data-Sci Bowl}}       \\ \hline
\multicolumn{1}{|c|}{\textbf{Methods}} & \multicolumn{1}{c|}{mDice}           & mIoU            & \multicolumn{1}{c|}{mDice}           & mIoU            & \multicolumn{1}{c|}{mDice}           & mIoU            & \multicolumn{1}{c|}{~~~~mDice~~~~}           & mIoU            \\ \hline
U-Net                                  & \multicolumn{1}{c|}{0.9145}          & 0.8654          & \multicolumn{1}{c|}{0.8629}          & 0.8176          & \multicolumn{1}{c|}{0.8554}          & 0.7847          & \multicolumn{1}{c|}{0.9080}          & 0.8314          \\ \hline
U-Net++                                & \multicolumn{1}{c|}{0.8453}          & 0.7559          & \multicolumn{1}{c|}{0.7475}          & 0.6313          & \multicolumn{1}{c|}{0.8094}          & 0.7288          & \multicolumn{1}{c|}{0.7705}          & 0.3010          \\ \hline
Deeplabv3+                             & \multicolumn{1}{c|}{0.8897}          & 0.8706          & \multicolumn{1}{c|}{0.8965}          & 0.8575          & \multicolumn{1}{c|}{0.8772}          & 0.8128          & \multicolumn{1}{c|}{0.8857}          & 0.8367          \\ \hline
MSRF-Net                               & \multicolumn{1}{c|}{0.9420}          & \textbf{0.9043} & \multicolumn{1}{c|}{0.9217}          & \textbf{0.8914} & \multicolumn{1}{c|}{0.8824}          & 0.8373          & \multicolumn{1}{c|}{0.9224}          & 0.8534          \\ \hline
\textbf{SSFormer-S}                    & \multicolumn{1}{c|}{0.9268}          & 0.8759          & \multicolumn{1}{c|}{0.9261}          & 0.8743          & \multicolumn{1}{c|}{0.9195}          & 0.8615          & \multicolumn{1}{c|}{\textbf{0.9254}} & \textbf{0.8652} \\ \hline
\textbf{SSFormer-L}                    & \multicolumn{1}{c|}{\textbf{0.9447}} & 0.8995          & \multicolumn{1}{c|}{\textbf{0.9357}} & 0.8905          & \multicolumn{1}{c|}{\textbf{0.9242}} & \textbf{0.8675} & \multicolumn{1}{c|}{0.9230}          & 0.8614          \\ \hline
\end{tabular}
\end{table}

\subsection{Results}

\subsubsection{Learning ability}
 We split the CVC-ClinicDB and Kvasir benchmark datasets into $80\%$ training set, $10\%$ evaluation set and $10\%$ test set according to the first scheme mentioned in Section \ref{Setup}. Table \ref{LA} demonstrates that our model improves the SOTA result by about $1.8\%$ on the CVC-ClinicDB benchmark and about $1\%$ on the Kvasir benchmark. These performances demonstrate the superior accurate prediction and learning abilities of SSFormer.

Furthermore, to assess the performance of SSFormer on other medical segmentation benchmarks, we conduct additional experiments on the ISIC-2018 and 2018 Data Science Bowl benchmark datasets. The results in Table \ref{LA} reveal that our model achieves the SOTA and excellent performance on two benchmarks, 2018 Data Science and ISIC-2018, respectively.
 ~\\~\\
\begin{minipage}{\textwidth}\scriptsize
\renewcommand\arraystretch{1.2}
        \begin{minipage}[t]{0.49\textwidth}
            \centering
            \makeatletter\def\@captype{table}\makeatother\caption{Generalization Test 1}
            \label{GT1}
            \begin{tabular}{|l|cc|cc|}
            \hline
\multicolumn{1}{|c|}{\textbf{Train Set}} & \multicolumn{2}{c|}{\textbf{CVC-ClinicDB}}               & \multicolumn{2}{c|}{\textbf{Kvasir-SEG}}             \\
\hline
\multicolumn{1}{|c|}{\textbf{Test set}} & \multicolumn{2}{c|}{\textbf{Kvasir-SEG}}               & \multicolumn{2}{c|}{\textbf{CVC-ClinicDB}}             \\ \hline
\multicolumn{1}{|c|}{\textbf{Methods}} & \multicolumn{1}{c|}{mDice}           & mIoU            & \multicolumn{1}{c|}{mDice}           & mIoU            \\ \hline
U-Net                                  & \multicolumn{1}{c|}{0.6222}          & 0.4588          & \multicolumn{1}{c|}{0.7172}          & 0.6133          \\ \hline
U-Net++                                & \multicolumn{1}{c|}{0.5926}          & 0.4564          & \multicolumn{1}{c|}{0.4265}          & 0.3345          \\ \hline
Deeplabv3+                             & \multicolumn{1}{c|}{0.6746}          & 0.5327          & \multicolumn{1}{c|}{0.6509}          & 0.5385          \\ \hline
MSRF-Net                               & \multicolumn{1}{c|}{0.7575}          & 0.6337          & \multicolumn{1}{c|}{0.7921}          & 0.6498          \\ \hline
\textbf{SSFormer-S}                    & \multicolumn{1}{c|}{0.7790}          & 0.6977          & \multicolumn{1}{c|}{0.7966}          & 0.7229          \\ \hline
\textbf{SSFormer-L}                    & \multicolumn{1}{c|}{\textbf{0.8270}} & \textbf{0.7348} & \multicolumn{1}{c|}{\textbf{0.8339}} & \textbf{0.7573} \\ \hline
\end{tabular}
        \end{minipage}
        \begin{minipage}[t]{0.531\textwidth}
        \centering
        \makeatletter\def\@captype{table}\makeatother
        \caption{Generalization Test 2}
        \label{GT2}
            \begin{tabular}{|l|cc|cc|}
\hline
\multicolumn{1}{|c|}{\textbf{Train Set}} &
\multicolumn{4}{|c|}{\textbf{Kvasir \& CVC-ClinicDB}}        \\
\hline
\multicolumn{1}{|c|}{\textbf{Test Set}} & \multicolumn{2}{c|}{\textbf{CVC-ColonDB}}            & \multicolumn{2}{c|}{\textbf{ETIS}}                   \\ \hline
\multicolumn{1}{|c|}{\textbf{Method}}  & \multicolumn{1}{c|}{\quad mDice \quad}          & mIoU           & \multicolumn{1}{c|}{mDic}           & mIoU           \\ \hline
UACANet-S                              & \multicolumn{1}{c|}{0.783}          & 0.704          & \multicolumn{1}{c|}{0.694}          & 0.615          \\ \hline
UACANet-L                              & \multicolumn{1}{c|}{0.751}          & 0.678          & \multicolumn{1}{c|}{0.766}          & 0.689          \\ \hline
CaraNet                                & \multicolumn{1}{c|}{0.773}          & 0.689          & \multicolumn{1}{c|}{0.747}          & 0.672          \\ \hline
PraNet                                 & \multicolumn{1}{c|}{0.712}          & 0.640          & \multicolumn{1}{c|}{0.628}          & 0.567          \\ \hline
\textbf{SSformer-S}                    & \multicolumn{1}{c|}{0.772}          & 0.697          & \multicolumn{1}{c|}{0.767}          & 0.698          \\ \hline
\textbf{SSformer-L}                    & \multicolumn{1}{c|}{\textbf{0.802}} & \textbf{0.721} & \multicolumn{1}{c|}{\textbf{0.796}} & \textbf{0.720} \\ \hline
\end{tabular}
        \end{minipage}
    \end{minipage}

\begin{figure}[htbp]
    \centering
    \includegraphics[width=12cm]{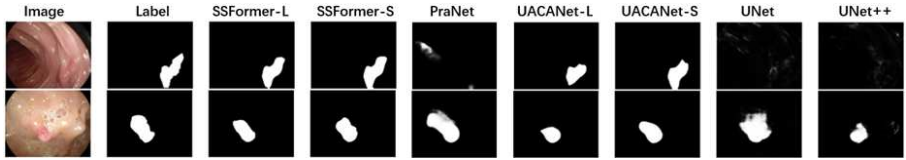}
    \caption{Predicted results of different methods}
    \label{Mask}
\end{figure}

\subsubsection{Generalization Ability}
We test the SSFormer trained on the CVC-ClinicDB and Kvasir datasets on the Kvasir and CVC-ClinicDB benchmarks, respectively. As mentioned in Section \ref{Setup}, this test result can reflect the generalization ability of our model. In Table \ref{GT1}, our model achieves outstanding performance using this testing scheme. In addition, to further assess the generalization ability of SSFormer, we refer to the experimental scheme of PraNet, use the training set constructed from part of the Kvasir and CVC-ClinicDB datasets for training, and test the model on the CVC-ColonDB and ETIS benchmarks. The results in Table \ref{GT2} demonstrate that our model significantly improves the SOTA performance ($3\%$) in the most challenging ETIS and achieves superior performance in CVC-ColonDB. Fig. \ref{Mask} gives the prediction accuracy of our model on the ETIS benchmark. These results can prove that SSFormer has robust generalization and accurate prediction abilities. (The scores in Table \ref{GT1} and Table \ref{GT2} are obtained from \cite{UACAnet,u-net,srivastava2021msrf,unet++})

\begin{table}[h]\scriptsize
\centering
\caption{Different encoder and decoder combinations performance with the same hyperparameter settings. The performance of different encoder and decoder combinations. The score is the performance of the model on the (CVC-ClinicDB, Kvasir) dataset group. (SeD is Segformer's Decoder, MiT is the Segformer's Encoder, and the CvT is proposed in \cite{cvt})}
\label{AS2}
\renewcommand\arraystretch{1.3}
\begin{tabular}{|c|c|c|c|c|}
\hline
\multicolumn{1}{|l|}{Encoder\textbackslash{}Decoder} & MLA\cite{zheng2021rethinking}          & SeD\cite{xie2021segformer}          & PLD-Cat      & PLD-Add      \\ \hline
CvT\cite{cvt}                                                  &  ~0.898, 0.912~  &  ~0.820, 0.889~  &  ~0.912, 0.923~  & -            \\ \hline
PvT\cite{pvtv1}                                                  &  0.809, 0.799  &  0.588, 0.618  &  0.828, 0.801  & -            \\ \hline
MiT\cite{xie2021segformer}                                                  &  0.907, 0.893  &  0.911, 0.903  &  0.916, 0.925  &  ~0.923, 0.897~  \\ \hline
\end{tabular}
\end{table}

\begin{table}[h]\scriptsize
\centering
\caption{The effect of PLD components on model performance. The score in the table follow (mDice, mIOU).}
\label{AS3}
\renewcommand\arraystretch{1.3}
\begin{tabular}{|r|c|c|c|}
\hline
\multicolumn{1}{|l|}{Decoder\textbackslash{}Dataset} & Kvasir-SEG            & ISIC-2018             & 2018 Data-Science Bowl \\ \hline
Without PLD                                          & 0.869, 0.918          & 0.855, 0.894          & 0.835, 0.904           \\ \hline
LE                                                   & 0.877, 0.925          & 0.860, 0.909          & 0.850, 0.915           \\ \hline
SFA                                                  & 0.885, 0.930          & 0.863, 0.918          & 0.858, 0.920           \\ \hline
\textbf{LE+SFA}                                      & \textbf{0.891, 0.936} & \textbf{0.868, 0.924} & \textbf{0.861, 0.923}  \\ \hline
\end{tabular}
\end{table}

\subsection{Ablation Study}\label{section:AS}

In Table \ref{AS2}, the PLD performs the best with the MiT. We believe that this is because the convolution operation inside MiT can maintain the consistency of the spatial information of the model. Furthermore, the experiments in Table \ref{AS3} demonstrate the effectiveness of the PLD and its components.

\section{Conclusions}
In this research, we propose a novel deep learning model SSFormer, with robust generalization and learning ability.  These are critical for polyp segmentation. Furthermore, we find that our model also demonstrates powerful learning ability in ISIC-2018 and 2018 Data Science Bowl benchmarks in additional experiments. We believe that the SSFormer has great potential to improve deep learning performance in other medical image segmentation tasks. Furthermore, experiments demonstrate that our proposed local feature emphasis module effectively constrains the attention dispersion of Transformers. Therefore, our research can be further used to optimize the Transformer backbone network for the general computer vision community and high generalizability medical applications.

\subsubsection{Acknowledgments}
This work was supported by the Key Program Special Fund in XJTLU (KSF-A-22).

\bibliography{ssformer}

\end{document}